\documentclass[useAMS]{mn2e}
\usepackage{graphicx}
\usepackage{multirow}
\usepackage{txfonts}
\usepackage{color}

\title[Pulsations in B-type supergiants with masses $M<20M_\odot$ before and after core helium ignition]{Pulsations in B-type supergiants with masses $M<20M_\odot$ before and after core helium ignition}
\author[J. Ostrowski and J. Daszy\'nska-Daszkiewicz]
{J. Ostrowski$^{1}$\thanks{E-mail: ostrowski@astro.uni.wroc.pl} and
J. Daszy\'nska-Daszkiewicz$^{1}$\thanks{E-mail: daszynska@astro.uni.wroc.pl}\\
$^{1}$Instytut Astronomiczny, Uniwersytet Wroc{\l}awski, ul. Kopernika 11, 51-622 Wroc{\l}aw, Poland\\}
\begin{document}

\date{Accepted 1988 December 15. Received 1988 December 14; in original form 1988 October 11}

\pagerange{\pageref{firstpage}--\pageref{lastpage}} \pubyear{2002}

\maketitle

\label{firstpage}

\begin{abstract}
Evolutionary tracks and pulsational analysis of models with masses of 13-18 $M_\odot$ are presented. We address two important questions. The first one deals with one of the most unresolved problems in astrophysics, i.e., the existence of a blue loop after core helium ignition; the so called "to loop or not to loop" problem. We show that inward overshooting from the outer convective zone in the red giant phase is prerequisite for the development of the blue loop. Our second question concerns pulsational instability of models in the core helium burning phase. We present for the first time that models on the blue loop can have unstable modes driven by the $\kappa$ mechanism operating in the $Z-$bump. Contrary to post-main sequence models in the shell hydrogen burning phases, pulsational instability
of the blue loop models depends mainly on effective temperature and metallicity is of secondary importance. Finally, we try to interpret the oscillation spectrum of the blue supergiant HD 163899, the only member of the SPBsg class, and to get some clue on the evolutionary status of the star.
\end{abstract}


\section{Introduction}

Slowly Pulsating B-type supergiants (SPBsg) have emerged as a new class of pulsating variables after Saio et al. (2006) have detected
pulsational frequencies in the MOST satellite data of the blue supergiant HD 163899 (B2 Ib/II, Klare \& Neckel 1977, Schmidt \& Carruthers 1996).
Saio et al. (2006) identified 48 frequency peaks in the MOST light curve and attributed them to g- and p-mode pulsations.
Because so far only one object of this type is found, the SPBsg class is not yet well defined as for the range of effective temperature and luminosity.
Also the evolutionary status of these pulsators remains unrevealed; they may be either in the shell hydrogen burning phase as well as after core helium ignition.

The finding of Saio et al. (2006) has prompted a few groups (Godart et al. 2009, Daszy\'nska-Daszkiewicz, Ostrowski, Pamyatnykh. 2013; hereafter D2013) to reanalyse pulsation stability in models of B-type stars after the Terminal Age Main Sequence (TAMS) and to further studies of SPBsg variables. The presence of g-mode pulsations in B-type post-main sequence stars has been explained by a partial reflection of some modes at an intermediate convective zone (ICZ) related to the hydrogen-burning shell or at a chemical gradient zone surrounding the radiative core.

However, all studies of these objects published so far are based on the assumption that HD 163899 has not reached the phase of core helium ignition, i.e., it is in the phase of shell hydrogen burning. This assumption might not necessary be fulfilled, because the blue loop can reach temperatures of early B spectral types. In this paper we calculate models which undergo core helium burning on the blue loop and compare them with models in the hydrogen shell burning phase.

The structure of the paper is as follows. In Section \ref{domains}, we present the new instability domains which include pulsational instability on the blue loops.
The effects of opacity, metallicity, overshooting, element diffusion and mass loss on the pulsational instability areas are studied.
Propagation diagrams and properties of instability parameter and kinetic energy of modes for representative models are described in Section \ref{models}. In Section \ref{identification} we construct photometric diagnostic diagrams for the B-type supergiant models and discuss a prospect for mode identification.
An attempt to interpret the oscillation spectrum of HD 163899 is presented in Section \ref{hd163899}. The last section contains Conclusions.

\section{Theoretical models} \label{domains}

\begin{figure*}
\begin{center}
 \includegraphics[clip,width=88mm,angle=0]{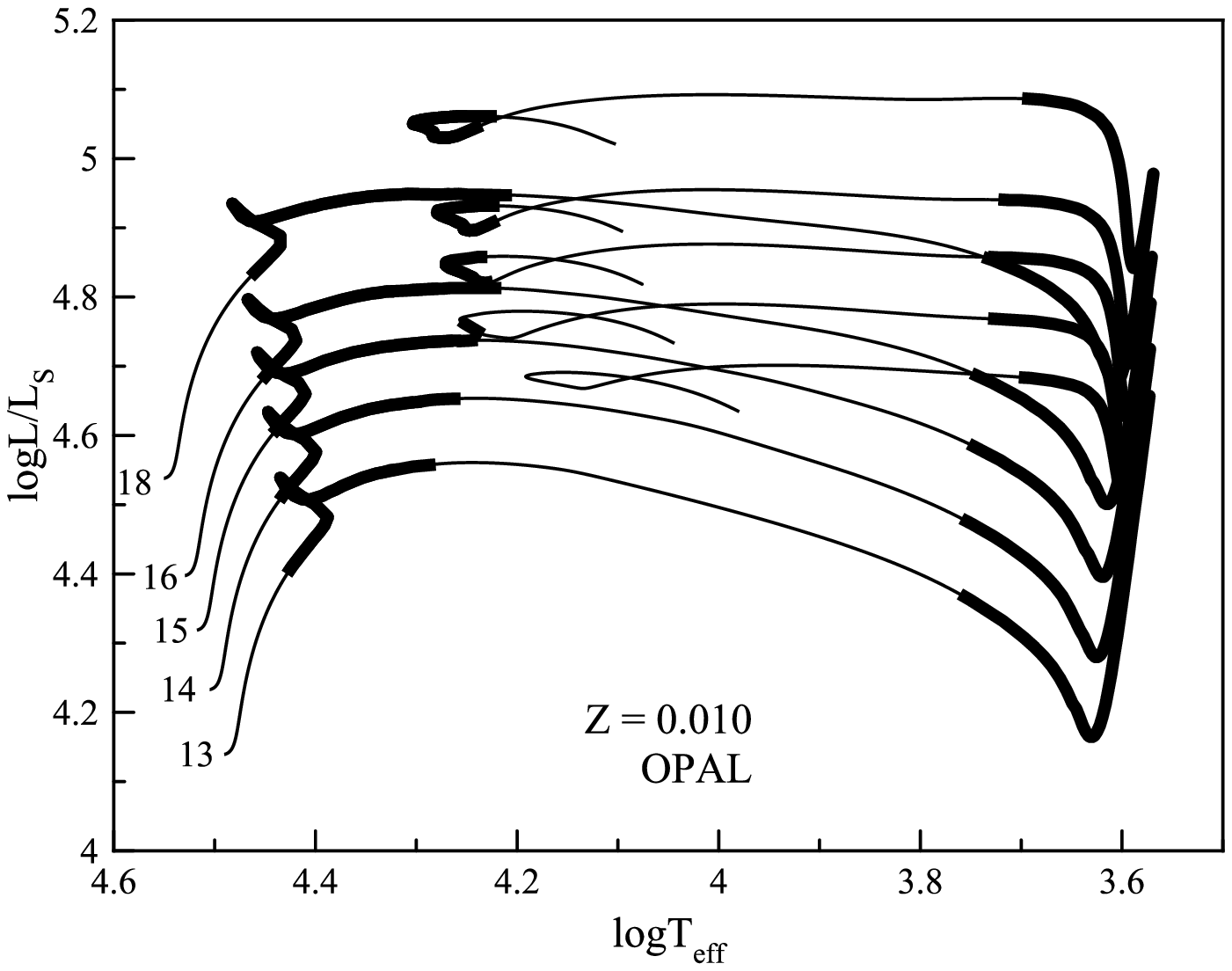}
 \includegraphics[clip,width=88mm,angle=0]{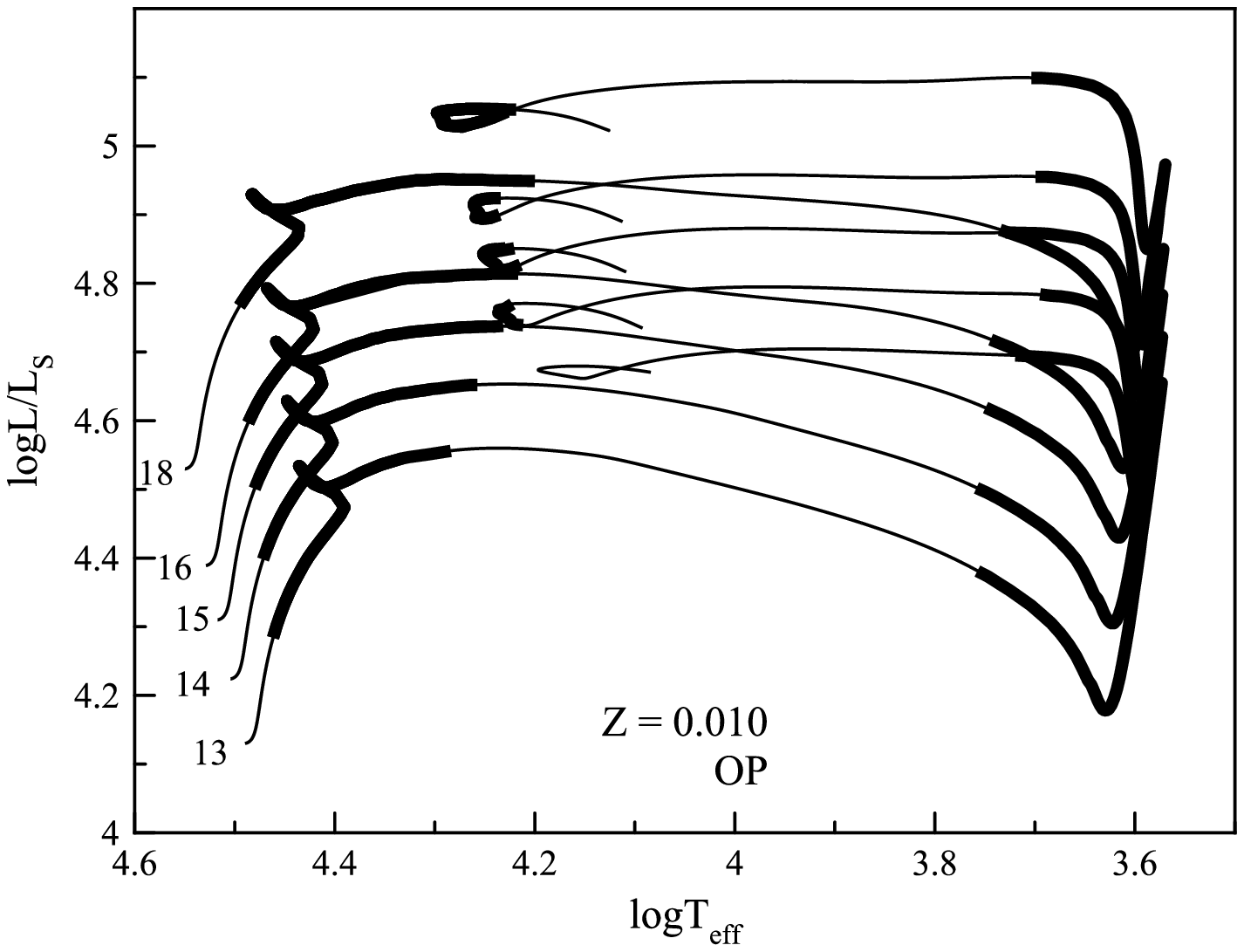}
 \includegraphics[clip,width=88mm,angle=0]{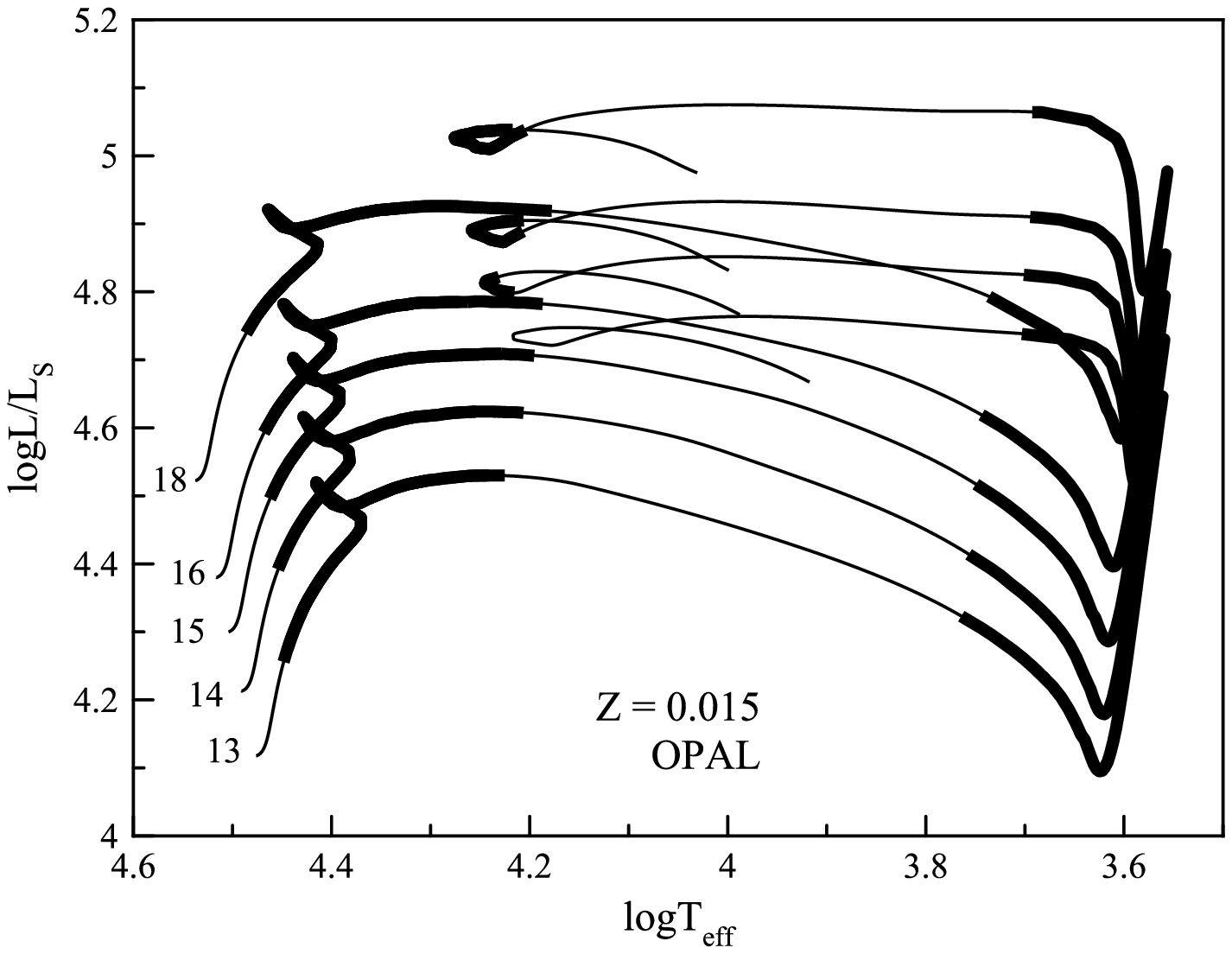}
 \includegraphics[clip,width=88mm,angle=0]{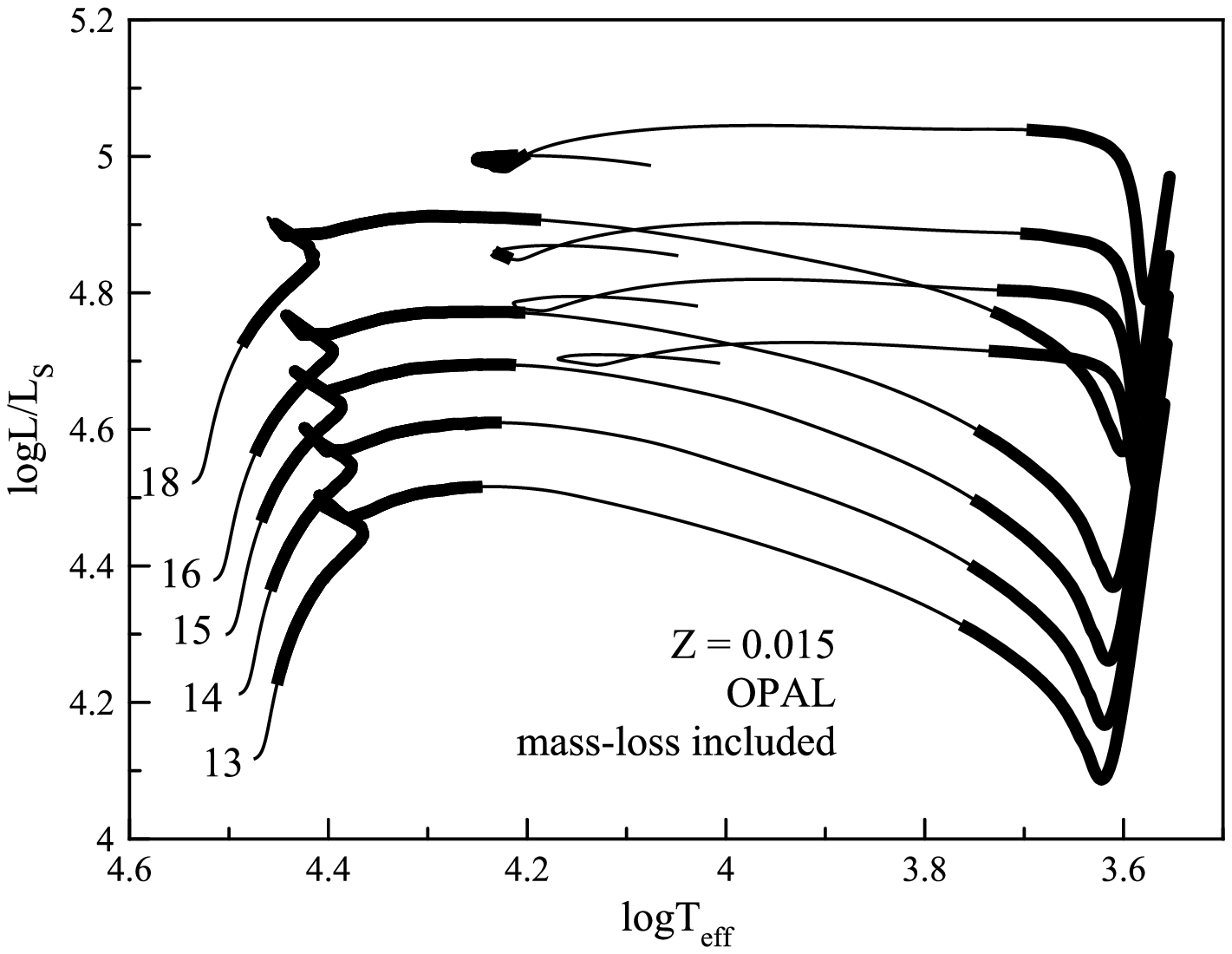}
 \includegraphics[clip,width=88mm,angle=0]{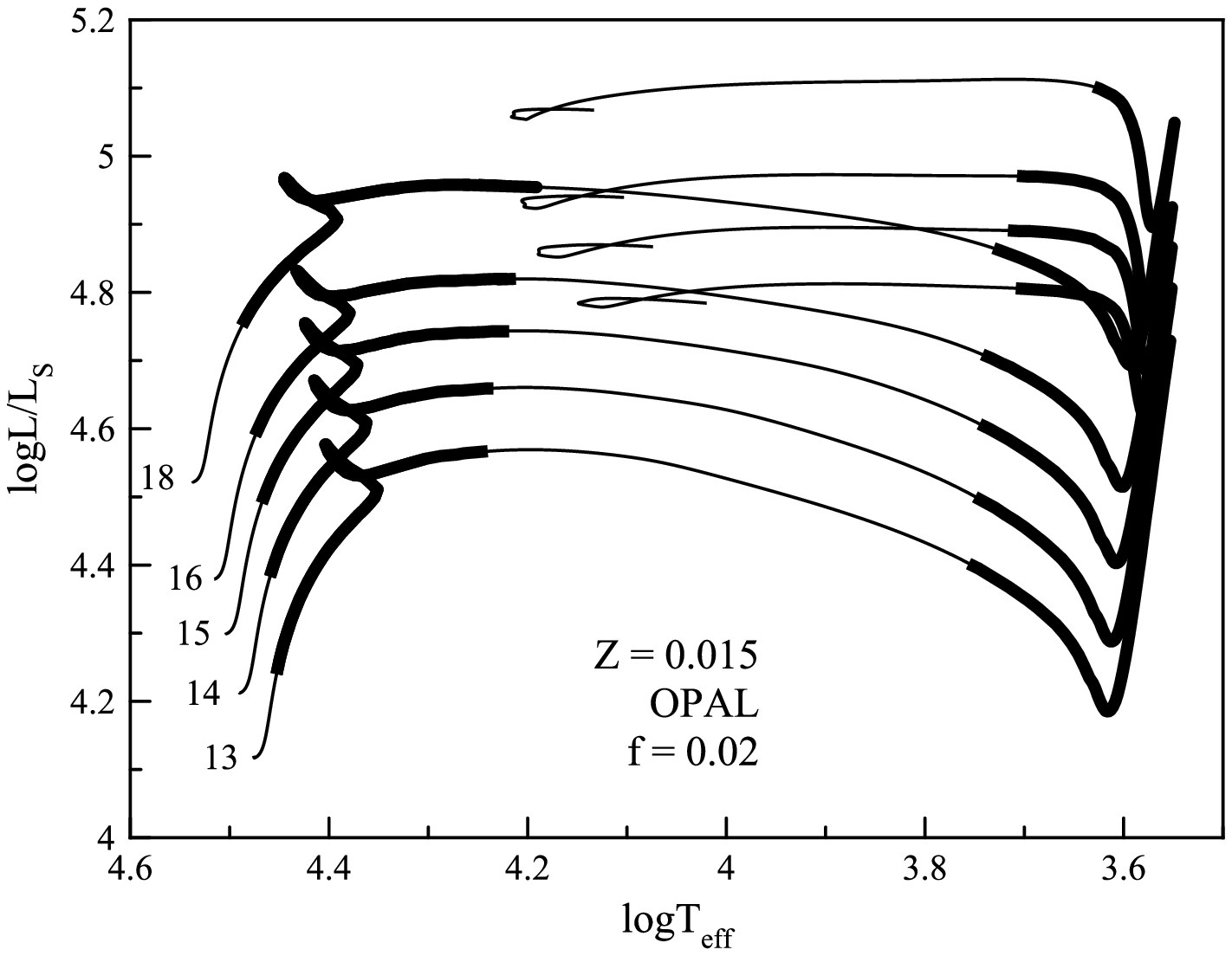}
 \includegraphics[clip,width=88mm,angle=0]{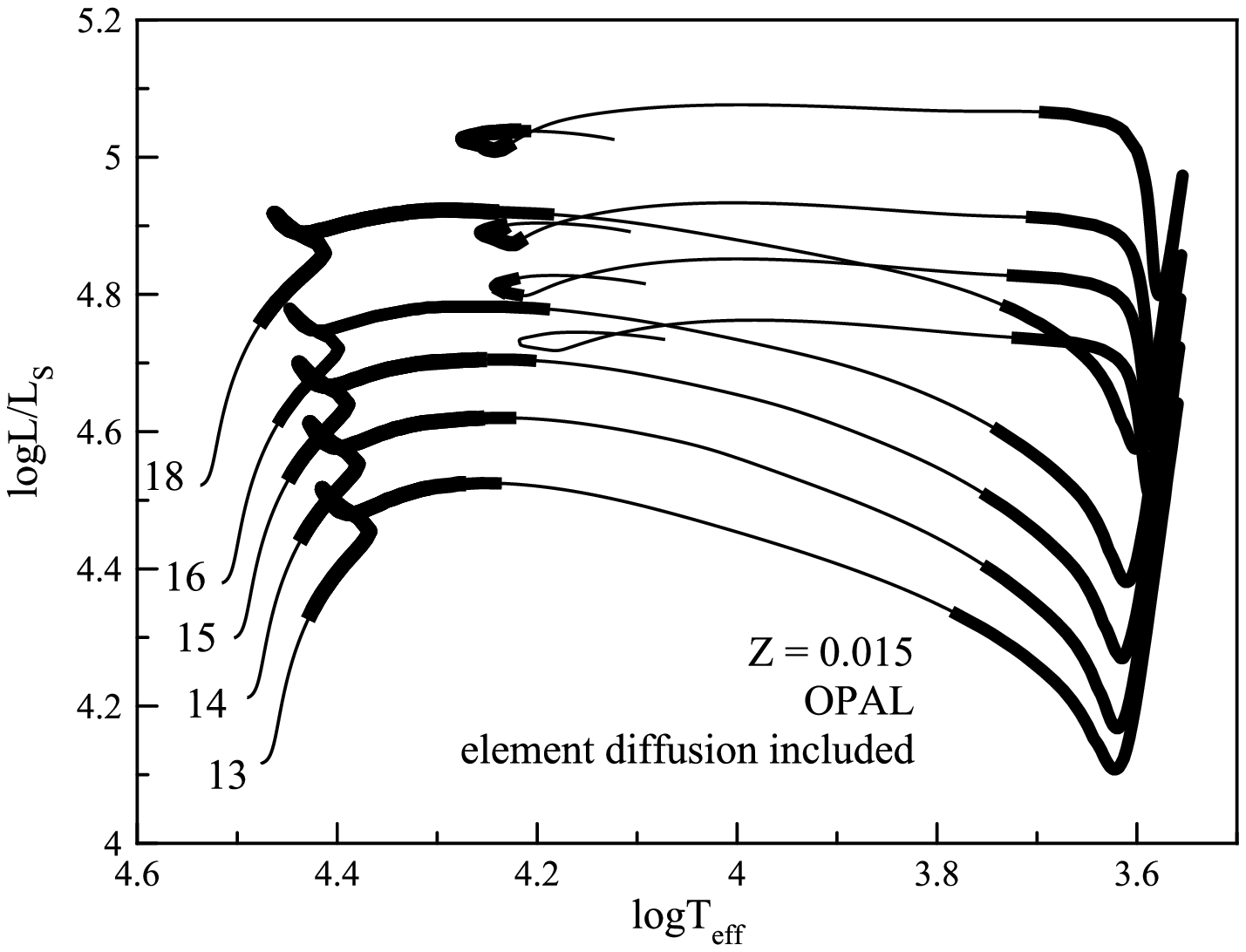}
   \caption{The HR diagram with evolutionary tracks for masses of $M = 13, 14, 15, 16$ and $18 M_\odot$ calculated until the end of the core helium burning. Two sources of opacity data were considered (OPAL and OP) for the two values of metallicty, $Z=0.01$ and $Z=0.015$, and two values of the overshooting parameter, $f=0.01$ and $f=0.02$. Various effects are shown. The top panels show models calculated for $Z=0.01$ and OPAL (top-left panel) and OP  opacity (top-right panel) tables. In the central panels we depict models calculated for $Z=0.015$ and OPAL tables without and with mass loss (central-left and central-right panels, respectively). The bottom-left panel shows the models calculated with $Z=0.015$, OPAL tables and convective overshooting parameter $f=0.02$. The bottom-right panel depicts models calculated with element diffusion.}
\label{fig1}
\end{center}
\end{figure*}

The evolutionary models were calculated with MESA evolution code (Modules for Experiments in Stellar Astrophysics, Paxton et al. 2011, Paxton et al. 2013). MESA allows to follow the evolution of massive stars to the pre-supernova phase. Our previous calculations (D2013) were limited to the phase before core helium ignition due to the lack of helium burning in Warsaw-New Jersey evolution code we used. We considered the mass range of 13$-$18 $M_\odot$. All computed models have an initial hydrogen abundance of $X=0.7$ and AGSS09 metal mixture (Asplund et al. 2009). The initial metal abundance varies from $Z=0.005$ to $Z=0.02$. OPAL (Iglesias \& Rogers 1996) and OP opacity tables (Seaton 2005) were used. All effects of rotation were neglected. Non-adiabatic pulsation analysis was performed using the code of Dziembowski (1977). In Fig.\,\ref{fig1} we present evolutionary tracks including various effects: opacity, metallicity, overshooting, mass loss and element diffusion.
They will be discussed in detail in the following subsections.

\subsection{Treatment of convection and the blue loop problem} \label{loops}

The convective zones are determined by the Ledoux criterion. It seems to be more justified than the Schwarzschild criterion because of a direct sensitivity of the chemical gradient to evolution. Lebreton et al. (2009) argued that the use of the Ledoux criterion leads to a narrower ICZ and hence to the smaller number of unstable modes. In case of our models, MESA produces the wide ICZ with both criterions and this effect is not significant. For the envelope, we adopted the parameter of the mixing length theory (MLT) of $\alpha_\mathrm{mlt}=1.8$ which is the higher value than usually used for massive stars. During the main sequence (MS) and hydrogen-shell burning phases, the value of  $\alpha_\mathrm{mlt}$ has a negligible effect on structure of massive stars but on the red giant branch (RGB) a higher efficiency of convection is needed and hence we opted for the higher value of $\alpha_\mathrm{mlt}$. We took into account convective overshooting from the hydrogen and helium core and inward overshooting from non-burning convective zones, using the exponential formula (Herwig 2000):
\begin{equation}
  \label{overshooting}
  D_\mathrm{OV} = D_\mathrm{conv}\exp(-\frac{2z}{fH_P}),
\end{equation}
\noindent where $D_\mathrm{conv}$ is the diffusion coefficient derived from MLT at a user-defined location near the boundary of a convective zone, $H_P$ is the pressure scale height at that location, $z$ is the distance in the radiative layer away from that location, and $f$ is an adjustable parameter, which we set to $0.01$ for most of calculated models. This value of $f$ mimics the behaviour of step overshooting with $\alpha_\mathrm{ov}\approx 0.1H_P$.

\begin{figure}
\begin{center}
 \includegraphics[clip,width=86mm]{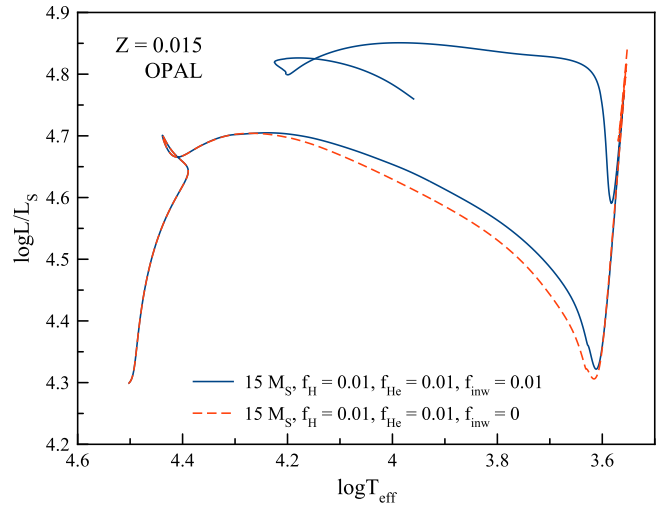}
   \caption{Comparison of evolution tracks calculated for models with mass $M = 15 M_\odot$ with (solid line) and without (dashed line) inward overshooting from the non-burning convective zone. OPAL opacity tables and metallicity $Z = 0.015$ were used.}
  \label{fig2}
\end{center}
\end{figure}

In our models the core helium ignition commences on the way towards the RGB but at this stage the energy produced from the helium burning is much lower than the energy produced from the hydrogen shell burning. The shell dominates the evolution up to the tip of RGB. At this point the energy produced in a helium core becomes comparable to the energy from the shell and the luminosity of a star drops. A star might eventually move towards higher effective temperatures and form a blue loop. The behaviour of the blue loops is still very poorly known and it is difficult to predict the effect of different parameters on their properties. We found that the convective overshooting has the most visible effect on their emergence, especially the inward overshooting from the non-burning zone. Without that overshooting it is impossible to obtain blue loops in our models independently of metallicity, opacity, diffusion, etc. This effect is shown in Fig.\,\ref{fig2}, where we depicted two evolutionary tracks for the mass of 15 $M_\odot$
calculated with and without inward overshooting from the non-burning zone. The blue loop emerges if the $\mu$-gradient left by the evolution during MS is erased by the outer convective zone on the RGB. The range of this convective zone is increased by the inward overshooting. On the other hand, the overshooting from the hydrogen core generally acts against emerging of the blue loops. We show the effect of different values of the overshooting parameter on the evolutionary tracks in the central-left panel ($f=0.01$) and in the bottom-left panel ($f=0.02$) of Fig.\,\ref{fig1}. In both cases the blue loops are present but for the lower value of $f$ they have a larger extension. If we increase the overshooting only from the hydrogen core to $f=0.02$ and we leave $f=0.01$ for the helium core overshooting and the inward overshooting from non-burning zone, the blue loops still emerge but their extension is lower. This is caused mainly by the reduced overshooting from the helium core which decreases the amount of helium available to burn and hence leads to smaller extension of the blue loop.

The presented examples show that properties of mixing, especially the convective overshooting, have a huge impact on the behaviour of the blue loops. It is difficult to discriminate between different effects but generally it seems that the inward overshooting from the outer convective zone on the RGB is indispensable to obtain blue loops for the models in the studied range of masses. The overshooting from the hydrogen core suspends the emerging of the blue loops and the overshooting from the helium core has a slight effect on their extension.

\subsection{Instability domains}

For the models considered in Section\,2.1, we computed pulsational instability for modes with the degrees  $\ell = 0, 1, 2$.
They are shown as thick lines in in Fig.\,\ref{fig1}.
The top panels show models calculated with $Z=0.01$ and OPAL (the top-left panel) and OP opacity tables (the top-right panel). The central panels depicts models calculated for $Z=0.015$ and OPAL tables without and with mass loss (the central-left and central-right panels, respectively). The bottom-left panel shows the models calculated with $Z=0.015$, OPAL tables and convective overshooting parameter $f=0.02$. The bottom-right panel depicts models calculated with element diffusion. Details of the used mass-loss models and implementation of diffusion are described later in this section. Similarly to our previous results (D2013) there is the p- and g-mode instability beyond TAMS for the models that undergo hydrogen shell burning. Most of the unstable modes are non-radial, but the radial fundamental mode is also unstable in hotter models beyond main sequence.

The main difference between our current and all previous calculations is the presence of unstable non-radial modes on the blue loops for more massive models ($M \gtrsim 14 M_\odot$ and $\log T_\mathrm{eff} \gtrsim 4.2$). The radial modes are stable in this evolutionary stage in all calculated models. The existence of pulsation instability on the blue loop, as well as the blue loops themselves, depend critically on the metallicity, $Z$. With lowering the metallicity the blue loops reach the higher effective temperatures which has a huge impact on pulsation instability. In models with the lower value of $\log T_\mathrm{eff}$ the driving layer related to the metal opacity bump is located too deep in the star to be effective. On the other hand, the lower value of metallicity the lower the $Z$ opacity bump. This crucially weakens instability for the MS and H-shell burning models but it seems to have a negligible effect on the instability on the blue loops, especially when compared to the effect of the effective temperature. In models with $Z \la 0.007$ the instability in the phase before core helium ignition is entirely quenched whereas there are still a lot of unstable modes in the models with very well developed blue loop. The pulsational instability on the blue loops vanishes at $Z\approx 0.004$. The reason for that is a difference in the internal structure resulting from a presence of an inward overshooting from the nonburning zone in the blue-loop models. The main effect of this overshooting is visible during the evolution
on the RGB, but there is also a thin zone of overshooting below the ICZ. That leads to enrichment of hydrogen in layers just below the ICZ, which causes a steep increase of the $\mu$-gradient in a thin zone. As a consequence the Brunt-V\"{a}is\"{a}l\"{a} frequency, $N^2$, increases as well, what makes trapping of pulsation modes more effective and some of them can be driven by the $\kappa$-mechanism operating in the Z-bump despite low metallicity.
The opposite situation occurs for models with higher metallicities. For example with $Z=0.02$ there is a large instability area beyond MS but there are no unstable modes in the blue loop models, because of too low temperature the blue loops reach. The exceptions are the highest considered masses - $M\approx 18 M_\odot$ for OPAL opacities and $M\approx 16 - 18 M_\odot$ for OP opacity tables. For those masses there are a few models on the hottest part of the blue loops that have unstable modes.

For every calculated mass and independently of metallicity or opacity tables we obtained unstable radial modes in the cooler models on the red giant branch or its vicinity. This result is similar to the one from Saio et al. (2013) and it was expected because in this area of HR diagram a classical instability strip is located. Those models are high-mass classical cepheids, in which the pulsations are driven in a region of the second ionization of helium. It is important to note that there are two effects that make the pulsational code we use unsuitable for such stars: the linear approximation and the lack of convection-pulsation interaction. We use
the convective flux freezing approximation which is not adequate in large outer convective zones. We do not study properties of these stars to any further extent in this paper.

\begin{figure}
\begin{center}
  \includegraphics[clip,width=88mm]{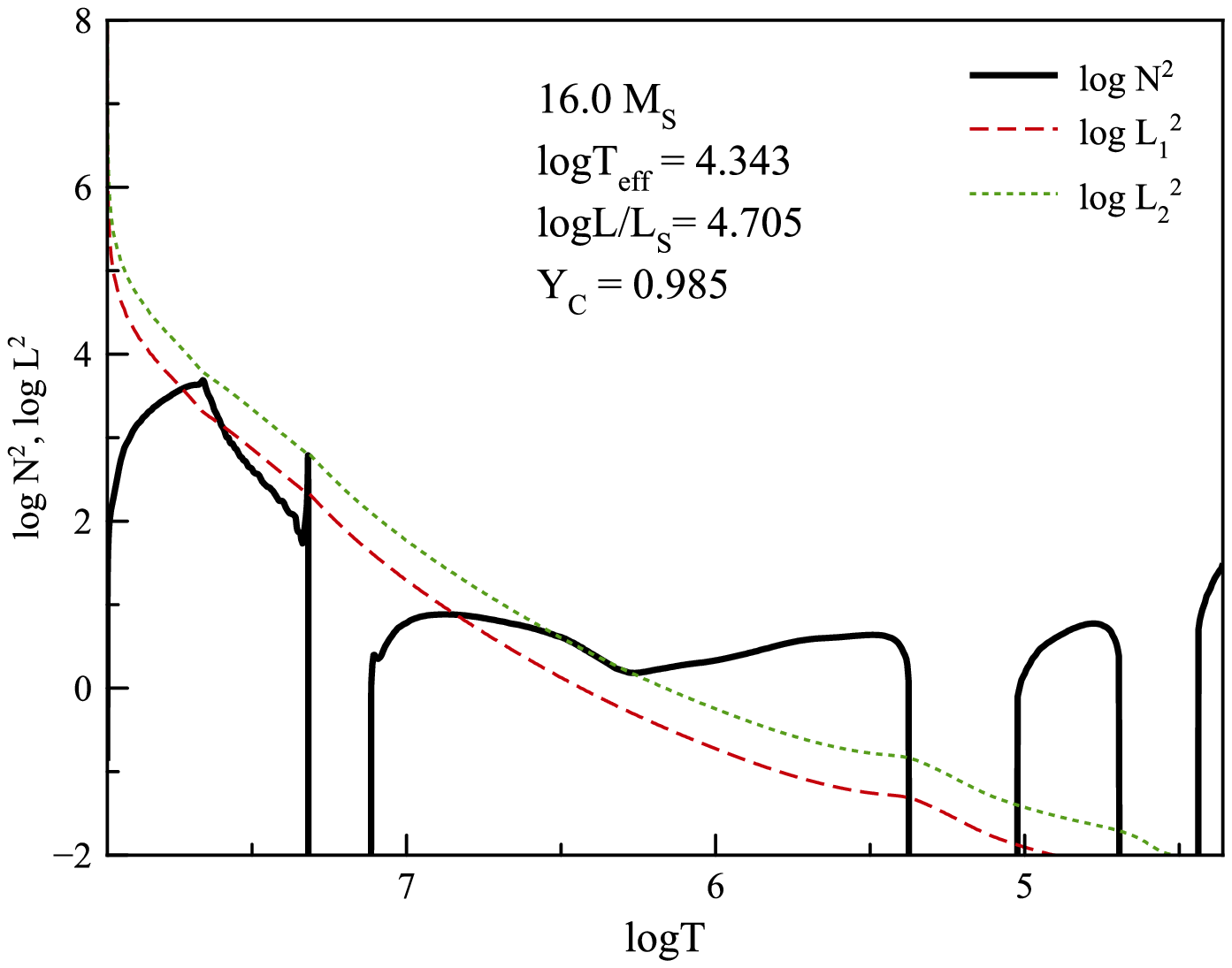}
  \includegraphics[clip,width=88mm]{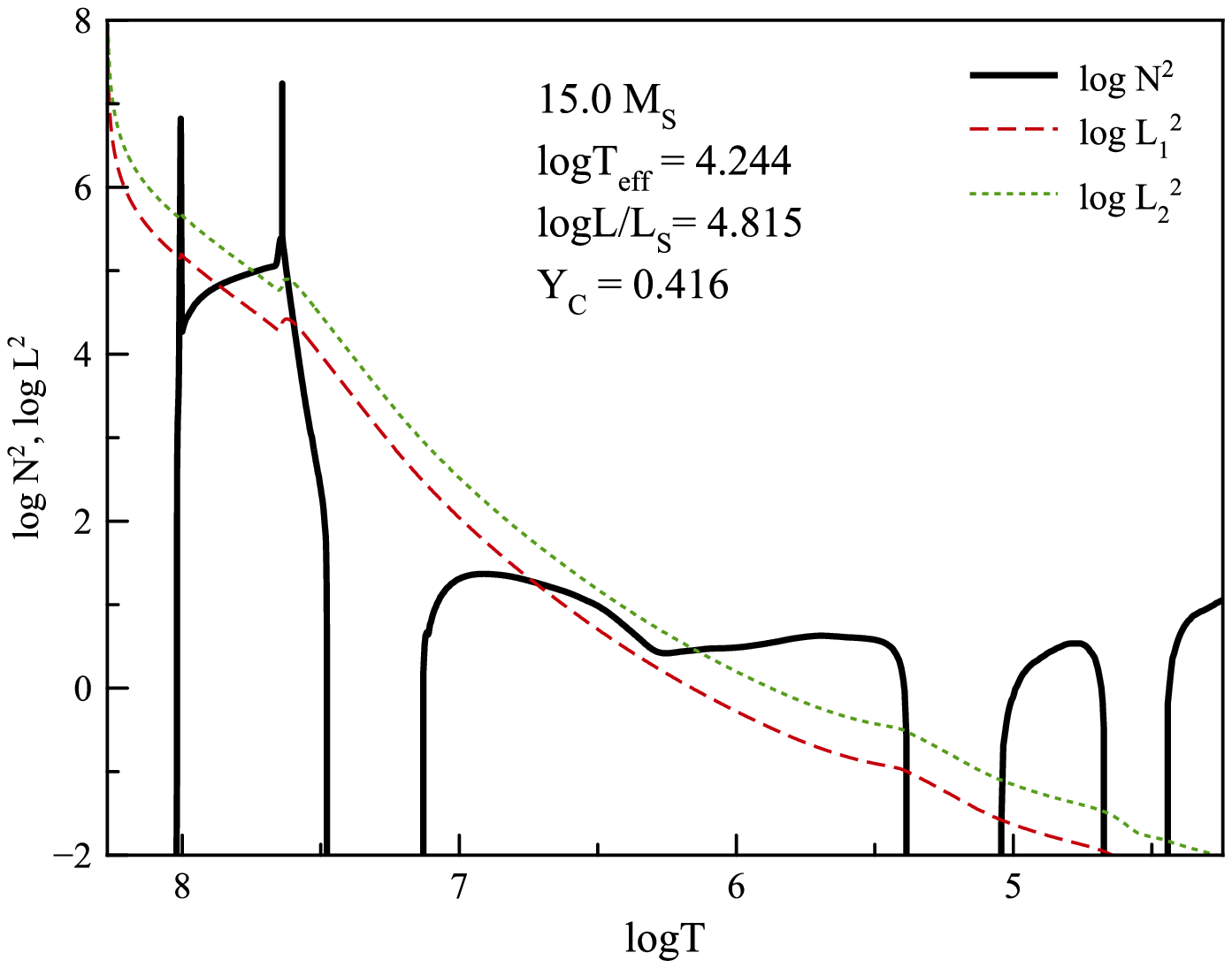}
   \caption{Runs of the Brunt-V\"{a}is\"{a}l\"{a} frequency, $\log N^2$ (solid lines) and and the Lamb frequencies, $\log L^2_\ell$, for $\ell=1$ (dashed lines) and $\ell=2$ (dotted lines), as a function of $\log T$, for Model\,1 (top panel) and Model\,2 (bottom panel). The scales of $\log T_\mathrm{eff}$ are different to represent the actual temperature profiles of the models.}
  \label{fig3}
\end{center}
\end{figure}

\begin{figure}
\begin{center}
 \includegraphics[clip,height=86mm, angle=-90]{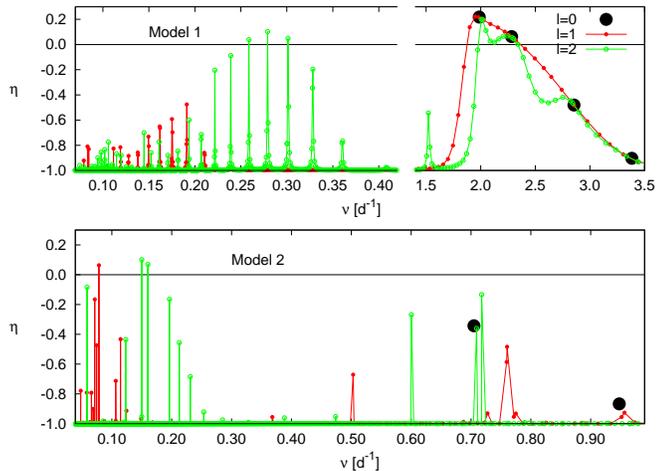}
   \caption{Instability parameter, $\eta$, as a function of frequency for $\ell=0,1,2$ for Model\,1 (top panel) and Model\,2 (bottom panel). Note a different ranges of frequency for both models.}
  \label{fig4}
\end{center}
\end{figure}

As depicted in Fig.\,\ref{fig1}, for $Z=0.01$ and $0.015$ there is a large instability strip that covers models both before and after core helium ignition. A comparison of the top- and central-left panels gives the effect of metallicity. For $Z=0.01$ (top-left panel), the range of effective temperatures that covers unstable models on the blue loop is a little wider than for $Z=0.015$ (central-left panel). This is a consequence of a larger extension of the loops towards the higher effective temperatures for lower metallicieties. As expected, an increase in $Z$ results in wider instability area on MS and beyond TAMS.

The upper panels show an effect of opacity tables on instability domains for $Z=0.01$, but the main conclusions are the same for every tested metallicity. There are significant differences on the MS phase, due to a long-known fact that for OP tables instability starts earlier than for OPAL tables. During the hydrogen shell burning phase and on the blue loop the differences are very subtle. The loops and instability areas have slightly different shapes but generally they are similar for both sources of the opacity data.

In the central panels we compare instability domains for models with $Z=0.015$ calculated without (the central-left panel) and with mass loss (the central-right panel). For $T_\mathrm{eff} > 10^4$ we used the wind's scheme of Vink et al. (2001) whereas for $T_\mathrm{eff} < 10^4$ a recipe of de Jager et al. (1988) was adopted. We do not study very massive models and hence the amount of mass which is lost during the evolution is rather modest. For instance, a star with initial mass of $16 M_\odot$ loses only about $0.8 M_\odot$ during the evolution from ZAMS to the point when helium is depleted in its core. The maximum value of the mass-loss rate is reached on the tip of the red giant branch and for the $16 M_\odot$ star it has the value of $\dot{M}_\mathrm{max}=10^{-5.6}~M_\odot/yr$. Mass loss seems to have a negligible effect on instability of models immediately beyond TAMS, whereas this effect on the instability on the blue loops is much more pronounced. With the reduced mass a star reaches lower luminosity during helium burning phase and also a blue loop reaches lower effective temperature than for models calculated without mass loss. Both of these effects lead to smaller instability range during this phase of evolution. The description of wind which we used for lower effective temperatures (de Jager et al. (1988)) gives relatively low values of mass loss when compared to other models we checked (Nieuwenhuijzen \& de Jager 1990, van Loon et al. 2005, Reimers 1975). It means that the instability areas on the blue loops calculated with other mass-loss models would be even smaller or entirely vanished. Also with sufficiently high mass loss the blue loops do not emerge at all. We will not discuss effects of mass loss anymore in this paper because they do not change any qualitative results presented here. Despite this it has to be remembered that it is a common phenomenon and it should be enabled during construction of detailed asteroseismic models.

The bottom-left panel depicts the instability areas calculated for higher value of convective overshooting, $f=0.02$. The most important difference when compared to models with $f=0.01$ (central-left panel) is smaller extension of the blue loops and hence there are no unstable models present during this evolutionary phase.

In the bottom-right panel of Fig.\,\ref{fig1} we present models calculated with $Z=0.015$, OPAL opacity tables and element diffusion. MESA calculates element diffusion by solving Burgler's equations using the method and diffusion coefficients of Thoul et al. (1994). The effect of diffusion on the shape of evolution tracks and instability areas is negligible, both for models before and after core helium ignition. The small changes of abundances of chemical elements in the central regions of the star do not modify significantly a behaviour of pulsation modes. The minor differences of frequencies might be important during a comparison to the observations but they do not influence the qualitative results presented here.

\section{Pulsation modes before and after core helium ignition} \label{models}

\begin{figure*}
\begin{center}
 \includegraphics[clip,width=88mm,angle=0]{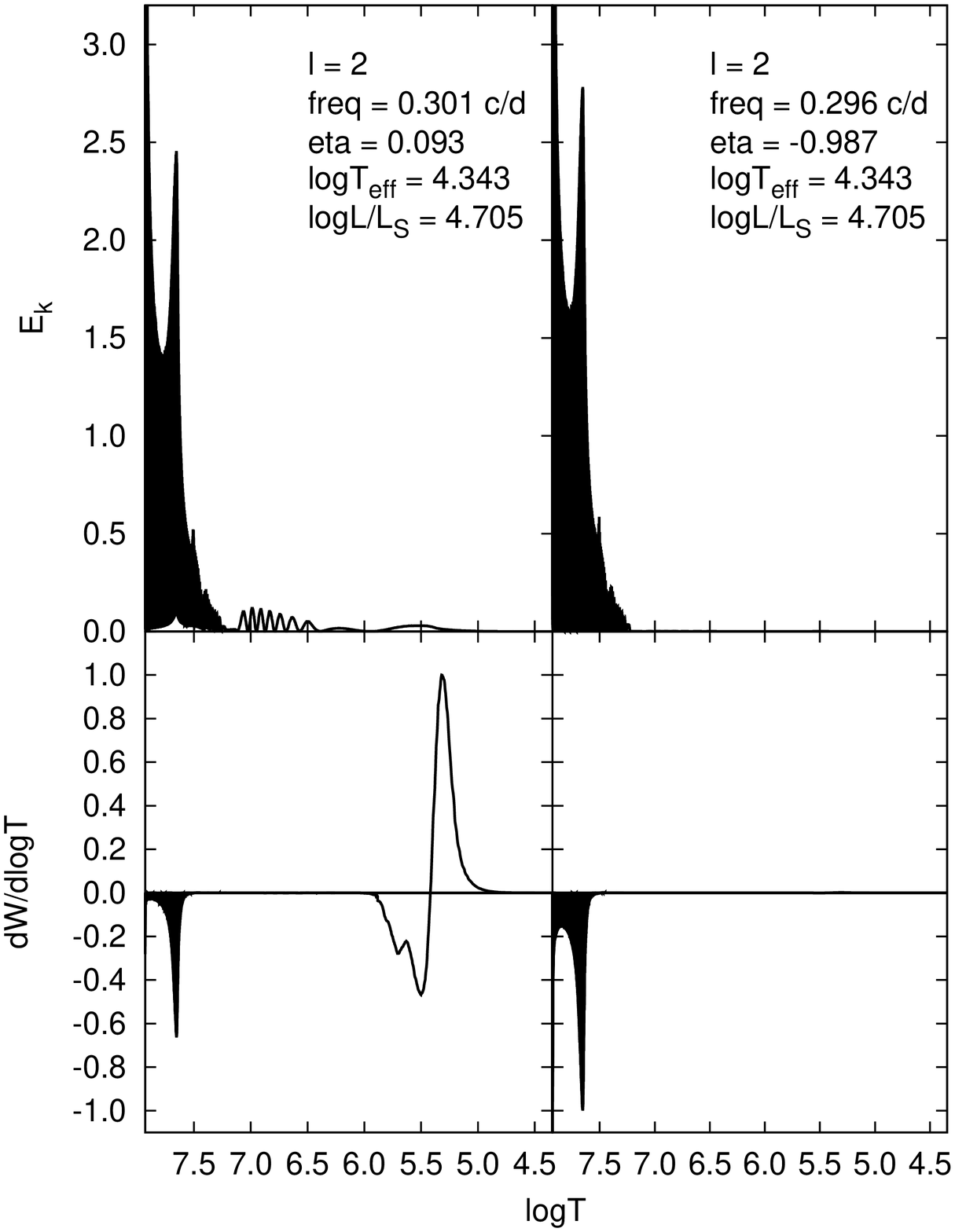}
 \includegraphics[clip,width=88mm,angle=0]{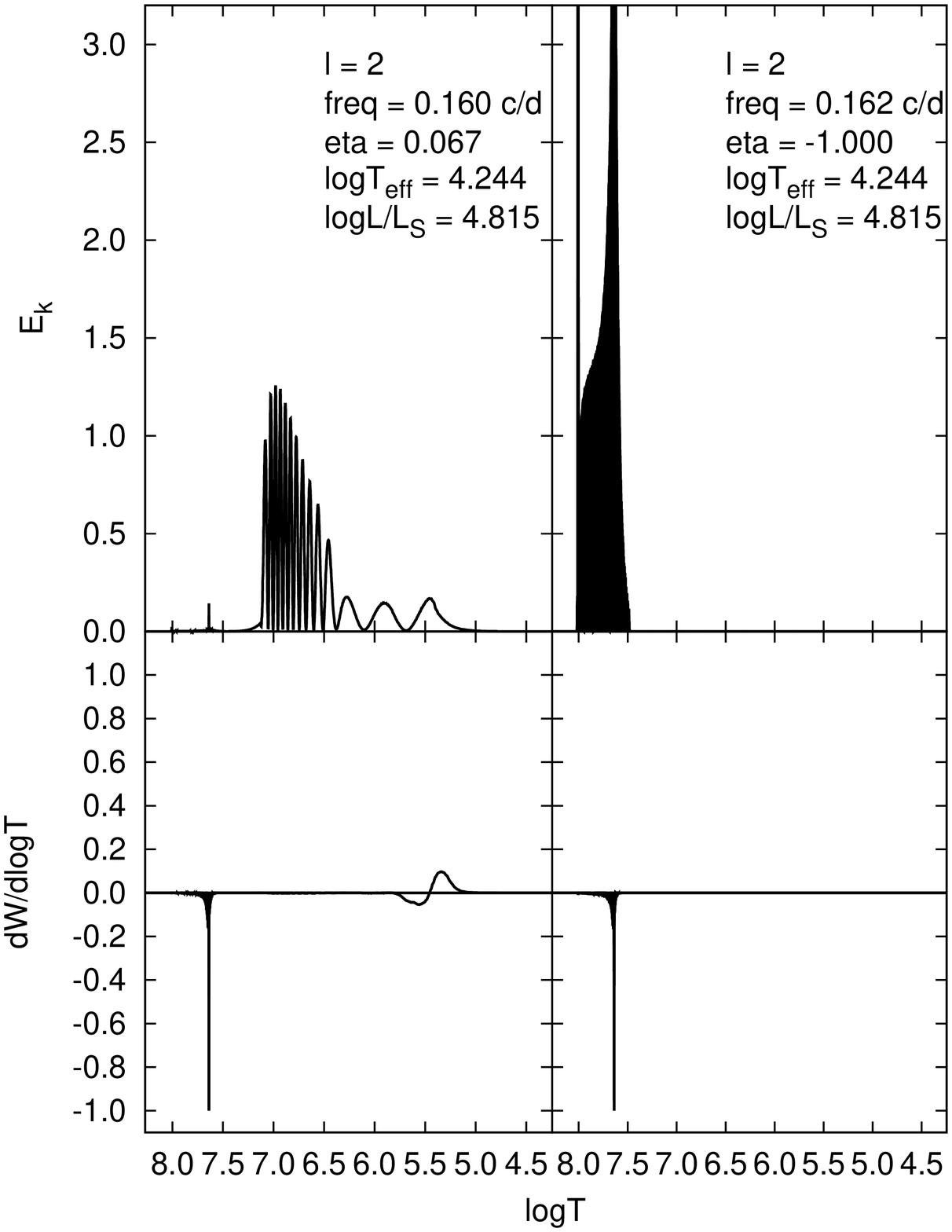}
   \caption{The kinetic energy density (top panels) and differential work integrals (bottom panels) for unstable and stable $\ell=2$ modes with close frequencies. The modes of Model\,1 and Model\,2 are presented in the left and right panels, respectively. The values of $\nu$ and $\eta$ are listed in panels.}
\label{fig5}
\end{center}
\end{figure*}

Here, we compare pulsation properties of two representative models with similar positions in the H-R diagram but in two different evolutionary stages, one during the shell hydrogen burning phase ($16 M_\odot$, $\log T_\mathrm{eff} = 4.343$, $\log L/L_\odot = 4.705$, hereafter Model\,1) and the second during core helium burning ($15 M_\odot$, $\log T_\mathrm{eff} = 4.244$, $\log L/L_\odot = 4.815$, hereafter Model\,2). They are calculated with OPAL opacity tables, AGSS09 metal mixture and $Z = 0.015$.

Propagation of pulsational modes in the radial direction is confined to the regions determined by the two characteristic frequencies (e.g., Unno et al. 1989), i.e., the Lamb (acoustic) frequency ($L_{\ell}$) and Brunt-V\"ais\"al\"a (buoyancy) frequency ($N$). The values of $L_{\ell}$ depend mostly on the mode degree, $\ell$, whereas the Brunt-V\"ais\"al\"a frequency is very sensitive to the structure of a star (e.g., a presence of convective zones or chemical gradients).

In Fig.\,\ref{fig3}, we present propagation diagrams, i.e., $L_\ell^2$, $N^2~~vs.~~\log T$, for Models 1 and 2 (the top and bottom panels, respectively). The main difference between these models is the presence of a convective core ($N^2 < 0$) in the model on the blue loop, whereas the model that undergoes shell hydrogen burning has a radiative core. Beyond the core, the propagation diagrams of these two models are qualitatively similar; in particular, both have a fully developed intermediate convective zone. The two spikes in the run of Brunt-V\"ais\"al\"a frequency that occur in the vicinity of convective zones in Model\,2 have different origin. The one at the border of the convective core is both a consequence of the very steep $\mu$-gradient in this region of star (the main reason) and also a numerical effect related to the extreme difficulties with precise determination of the border of helium core by evolution codes. Due to a very high sensitivity of triple-alpha process to a change of temperature, the location of border changes slightly with every step of evolution. That results in an artificial increase of the $\mu$-gradient, which is clearly visible in behaviour of $N^2$. The second spike, at the innermost border of ICZ is related to the presence of inward overshooting from this convective zone. That leads to a steep change of $\mu$ and a peak in the propagation diagram.

In Fig.\,\ref{fig4}, we present the instability parameter, $\eta$, for modes with the harmonic spherical degrees $\ell = 0,1,2$, as a function of frequency for Models 1 and 2 (the top and bottom panels respectively). Instability parameter measures the net energy gained by a mode during one pulsational cycle and is defined as (Stellingwerf 1978):

\begin{equation}
  \label{eta}
  \eta=\frac{W}{\int\limits_0^R\left|\frac{\mathrm{d}W}{\mathrm{d}r}\right|\mathrm{d}r},
\end{equation}

\noindent where $W$ is the global work integral. This definition gives normalization $\eta \in [-1,+1]$. If $\eta > 0$ a pulsation mode is unstable and if $\eta \le 0$ it is stable. There are two global maxima of $\eta$ in Model\,1, which is similar to the behaviour during the MS evolution (D2013). The first one, at low frequencies, is related to the very high-order g-modes, while the second one, at higher frequencies, is related to the low-order p- and g-modes. All non-radial modes in the second group are mixed modes and they behave like p-modes only in the outer envelope of the star. 
In the frequency range presented in Fig.\,\ref{fig4}, i.e., from 0.04 to 3.5 d$^{-1}$, the values of the radial order, $n$,
are in ranges of about $(20,~1700)$ for the $\ell=1$ modes and $n \in (30, 1650)$ for the $\ell=2$ modes.
In Model\,1 the fundamental radial mode and the first overtone are unstable with the ratio of periods $P_2/P_1 \simeq 0.87$. The most interesting behaviour of $\eta$ is visible in the vicinity of the first maximum. There is a clearly visible pattern of consecutive local minima and maxima, which does not occur in MS models. This phenomenon is related to the partial trapping of pulsation modes in the radiative core and the envelope. The detailed description of this phenomenon can be found in our previous paper (D2013).

In Model\,2, which burns helium in its core, a situation is different. There are only a few very high-order low-frequency g-modes which are unstable. These maximum values of $\eta$ are shifted towards lower frequencies when compared to Model\,1.
There is no pattern of local minima and maxima and the calculated frequency spectrum is much sparser than for Model\,1. This is typical for every blue-loop that a number of unstable modes is very low. Model\,2 has one unstable dipole mode with frequencies of $\nu\approx 0.08$ c/d and two unstable quadruple modes with frequencies $\nu\approx 0.14,~0.15$ c/d. Radial modes are always stable in models during core helium burning phase, regardless of metallicity, opacity tables, etc. However, the ratio of periods between the first overtone and the fundamental mode is visibly lower than in Model\,1 and has a value $P_2/P_1 \simeq 0.74$. In Model\,2, the radial orders of pulsation modes are much larger,in particular for the quadruple mode. For the considered range of $\nu$, cf. Fig.\,\ref{fig4}, the values of $n$ are: $(60,1400)$ for $\ell=1$ and $(100,2400)$ for $\ell=2$.

In Fig.\,\ref{fig5}, we compare the kinetic energy densities, $E_k$ (top panels), and differential work integrals, $\mathrm{d}W/\mathrm{d}\log T$ (bottom panels), for unstable and stable modes of the degree $\ell=2$ with very close frequencies. The left group of panels is for Model\,1 (before core helium ignition) and the left group of panels for Model\,2 (after core helium ignition). The behaviour of presented modes is typical and representative for all unstable and stable modes during shell hydrogen burning and on the blue loops. The properties of pulsational modes in the shell hydrogen burning model have been already presented in D2013 but for comparative purposes we repeat them here.

In Model\,1, both unstable ($\nu=0.301$ c/d, $\eta=0.093$) and stable ($\nu=0.296~d^{-1}$, $\eta=-0.983$) modes exhibit a very strong damping in the radiative helium core, which is visible in the behaviour of $\mathrm{d}W/\mathrm{d}\log T$ in the bottom panels of the left group. As shown in the upper panels of this group, most of the kinetic energy of the modes is confined in the central region. The most notable difference between the presented modes is that the unstable one has some energy deposited also in the radiative envelope, whereas the stable one has almost no energy in this region. This is due to the fact that the unstable mode has been partially reflected at the ICZ and is trapped in two cavities: the core and the outer layers. That energy, although its amount is small when compared to the amount trapped in the core, is sufficient for the $\kappa$ mechanism, operating in the $Z-$bump, to efficiently drive the pulsations. This is visible in the behaviour of differential work integral - there is a maximum around $\log T=5.2$ and it dominates over the damping in the core. On the contrary, the stable mode has not been reflected at ICZ because it does not have a node of eigenfunction there. That means that it penetrates the core with high amplitude and is effectively damped there. The above behaviour of $E_k$ explains the pattern of local maxima and minima of $\eta$ described earlier on.

For Model\,2 we also compare two very close modes: the unstable one with $\nu=0.160$ c/d and $\eta=0.067$ and the stable one with $\nu=0.162$ c/d and $\eta=-1.000$. Their kinetic energy densities and differential work integrals are shown in the right group of panels in Fig.\,\ref{fig5}. The behavior of $E_k$ is different for these modes than for the modes from the Model\,1. The entire energy of the stable mode is confined to the radiative zone in-between the convective core and the ICZ (there is strong damping in this region of the star), whereas the unstable mode has almost all of its energy trapped in the outer radiative zone. That means that the unstable modes of the blue loop models have to be almost entirely reflected at the ICZ. This fact might be the explanation why we do have far fewer unstable modes from models burning helium in its core than from models before core helium ignition and also the lack of the structure of $\eta$ observed in Model\,1. The behaviour of kinetic energy density depicted in Fig.\,\ref{fig5} is typical for all unstable and stable modes both during hydrogen shell burning and on the blue loops.

\section{Photometric amplitudes and phases of pulsations in early B-type supergiants} \label{identification}

\begin{figure*}
\begin{center}
 \includegraphics[clip,width=88mm,angle=0]{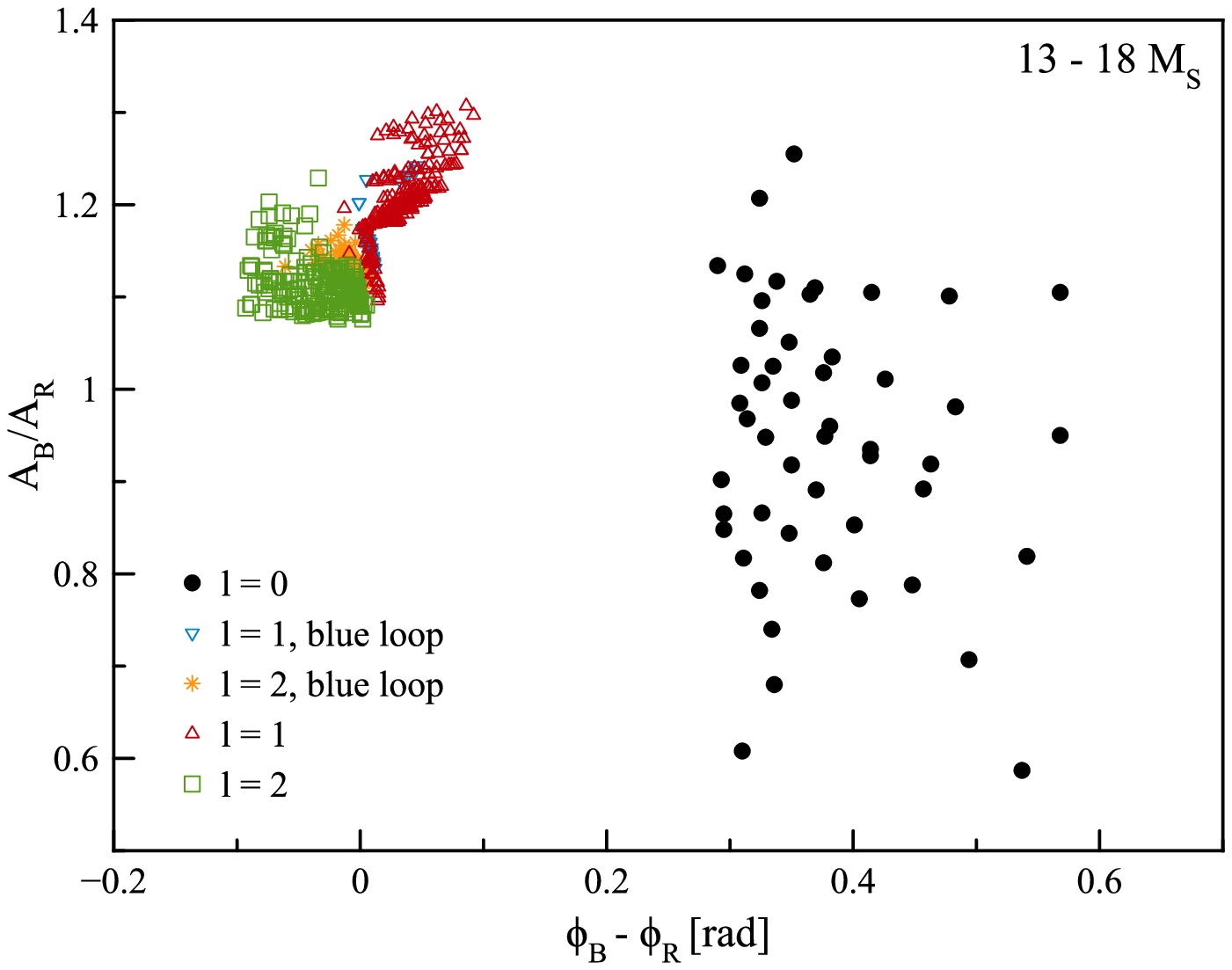}
 \includegraphics[clip,width=88mm,angle=0]{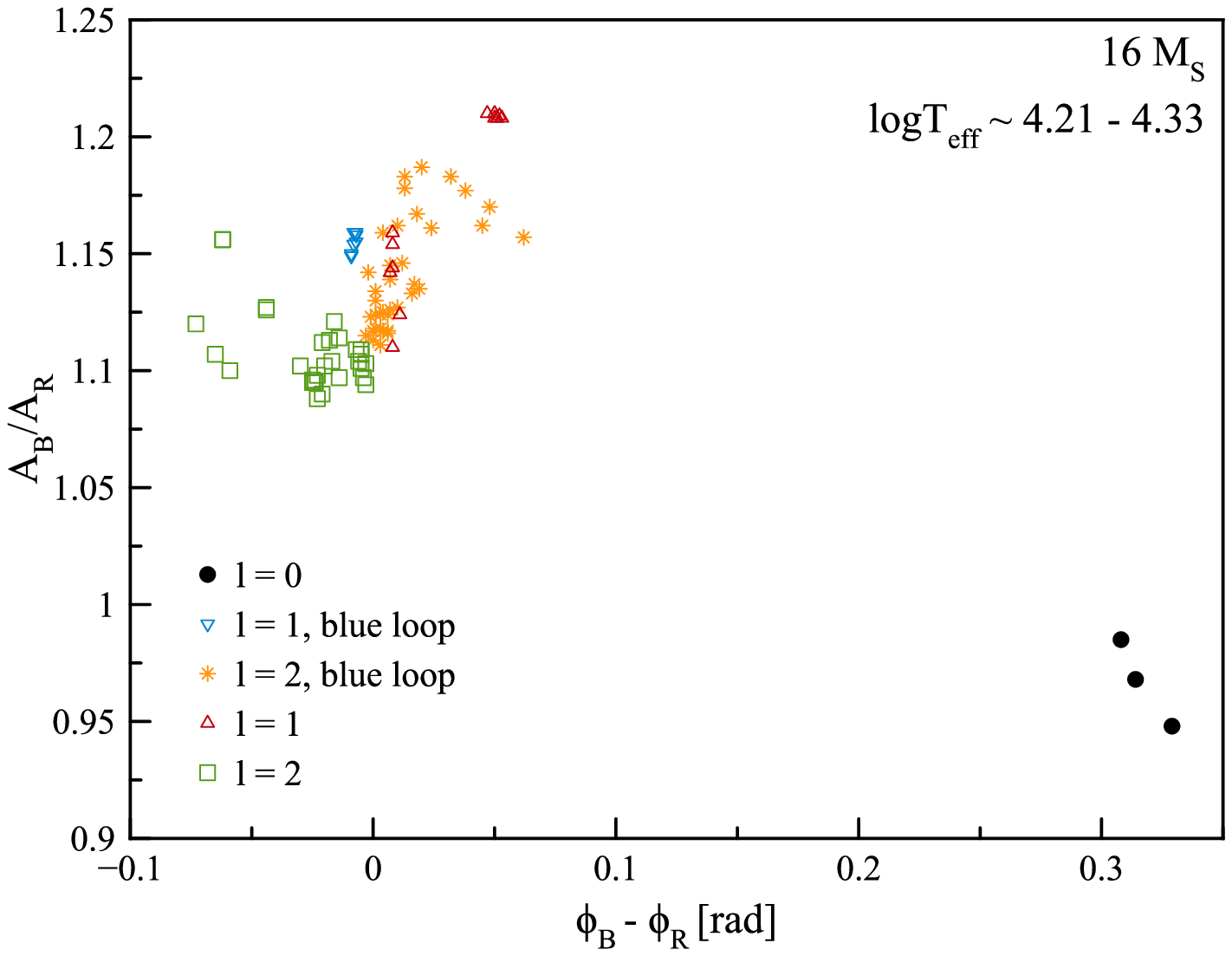}
   \caption{The photometric diagnostic diagrams in the Johnson BR passbands with positions of unstable modes with $\ell \le 2$ found in the post main sequence models, both before and after core helium ignition. The left panel includes unstable modes of the post-MS models with masses of 13, 14, 15, 16 and 18 $M_\odot$. The right panel includes unstable modes of the post-MS models with a mass of 16 $M_{\odot}$ and temperature in the range $4.21 < \log T_\mathrm{eff} < 4.33$. The OPAL opacities, $Z=0.015$ and NLTE model atmospheres were adopted.}
\label{fig6}
\end{center}
\end{figure*}

\begin{figure*}
\begin{center}
 \includegraphics[clip,width=85mm,angle=0]{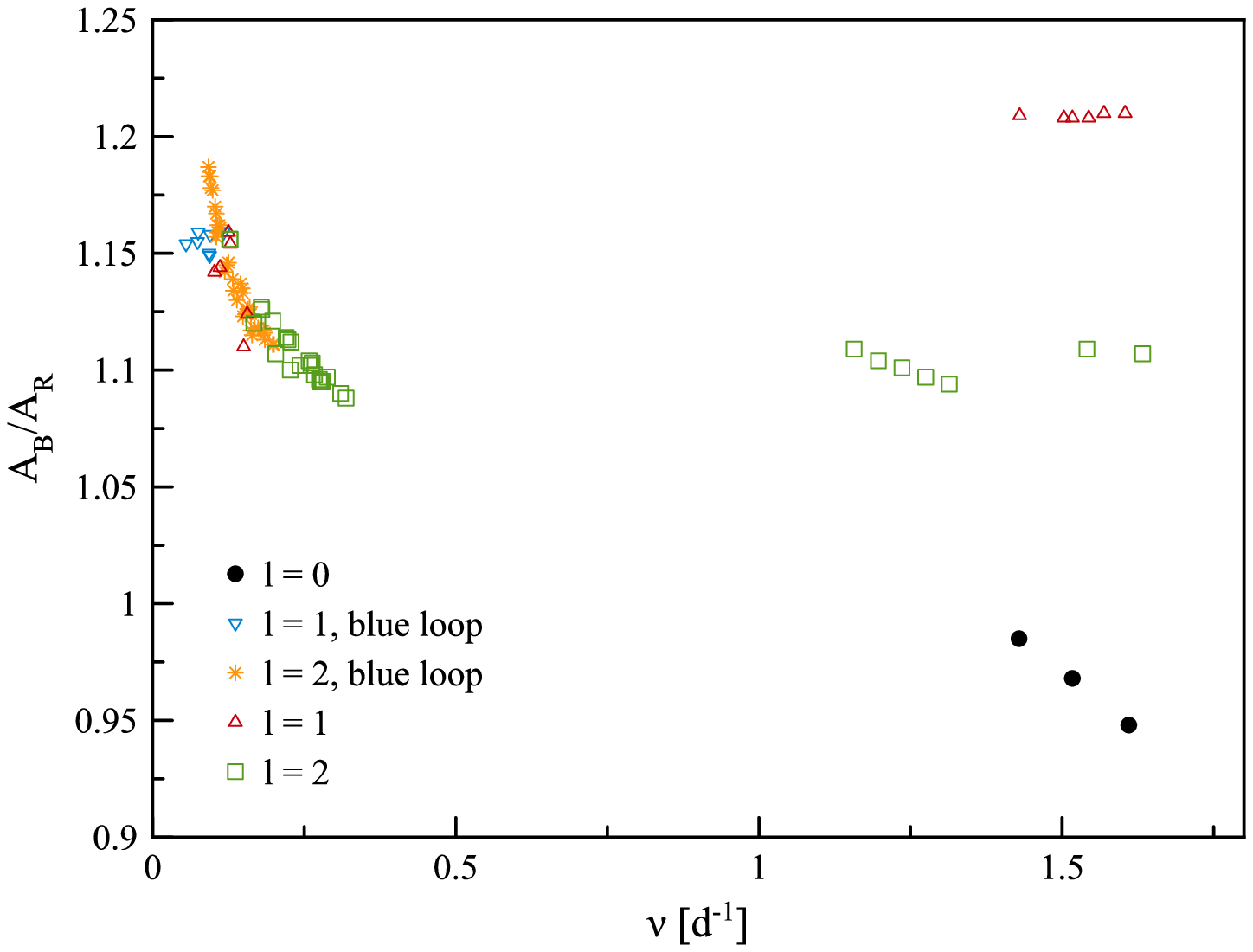}
 \includegraphics[clip,width=85mm,angle=0]{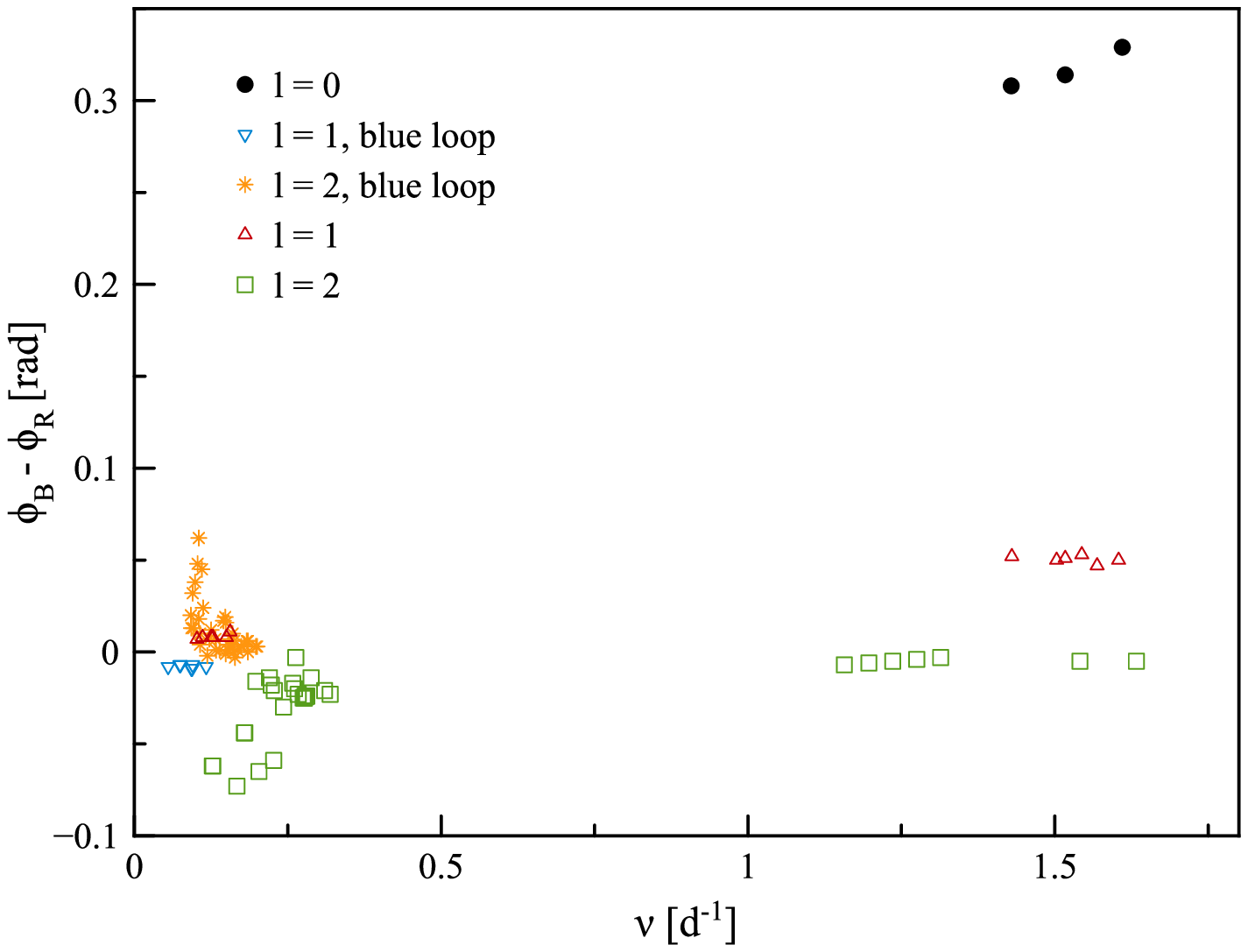}
   \caption{Amplitudes ratios (the left panel) and phase differences (the right panel) in the Johnson BR passbands, as a function of frequency, $\nu$, for unstable modes of the same models as considered in the right panel of Fig.\,\ref{fig6}.}
\label{fig7}
\end{center}
\end{figure*}

In our previous paper (D2013), we have shown that one should not expect regular patterns in the oscillation spectra of blue supergiants which the asymptotic theory predicts for g-mode pulsations. However, it turned out that there is a prospect for identification of the degree, $\ell$, for modes excited in the blue supergiant models before core helium ignition from photometric diagnostic diagrams (D2013). Here, we will check whether the photometric amplitudes and phases of pulsations in models after core helium ignition have similar properties.

The expression for the complex amplitudes of the light variations in a given passband  within the linear and zero-rotation approximation can be found, e.g., in Daszy\'nska-Daszkiewicz et al. (2002), and need not be repeated here. To compute the values of these observables, we used the nonadiabatic pulsational code of Dziembowski (1977) and tables of the fluxes in photometric passbands and nonlinear limb-darkening coefficients computed by Daszy\'nska-Daszkiewicz \& Szewczuk (2011) on the basis of the NLTE atmosphere models of Lanz \& Hubeny (2007).

In Fig.\,\ref{fig6}, we show positions of unstable modes with the degree $\ell=0-2$ excited in the supergiant models in the photometric diagnostic diagrams employing the $B$ and $R$ Johnson filters. These passbands are similar to those used on the BRIght-star Target Explorer Constellation (BRITE) which is already operating. We used the OPAL data for the AGSS09 mixture and $Z=0.015$. In the left panel we marked positions of all unstable modes in models with masses $13-18M_\odot$  whereas in the right panel we chose modes excited in models with  $16M_\odot$ and $4.21 < \log T_\mathrm{eff} < 4.33$.  These temperature range is a typical maximum error and larger than the error for HD 163899 (Saio et al. 2006). With different symbols we discriminate between different values of $\ell$ and evolutionary stages. Their meanings are described in the legend of the plots.

In the diagram for all studied masses, there is a perfect separation between radial and non-radial modes and fairly good separation between dipoles and quadruples. A discrimination between models undergoing shell hydrogen burning and models on the blue loops is not obvious. Modes excited in models on these evolution stages tend to group in slightly different locations but in a plot including such a range of masses their distinction would be uncertain given observational errors in the photometric amplitudes and phases. The situation improves slightly when we are able to constrain the mass and effective temperature. After fixing the range of $\log T_\mathrm{eff}$ in the right panel of Fig.\,\ref{fig6}, the separation between phases before and after core helium ignition can be noticeable.

Fig.\,\ref{fig7} presents how the photometric observables depend on the mode frequency considering again the BR Johnson passbands and
the same models as in the right panel of Fig.\,\ref{fig6}. From this plot we can conclude that much better separation between modes of different values of $\ell$,
from both the amplitude ratios and phase differences, occurs for higher frequency modes which are unstable only in Model\,1.


\section{Pulsations in HD 163899} \label{hd163899}

\begin{figure}
\begin{center}
 \includegraphics[clip,width=88mm,angle=0]{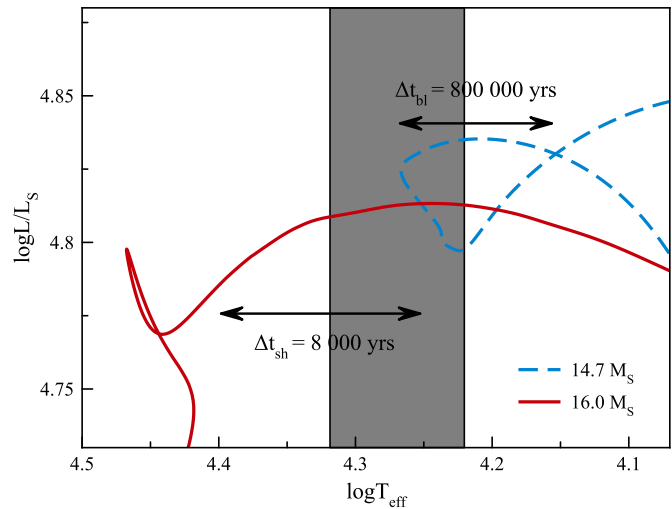}
   \caption{Comparison of evolution times in the pulsational unstable region of HR diagram during hydrogen shell burning phase and on the blue loop. The solid and dashed lines depict models with masses $16.0 M_\odot$ and $14.7M_\odot$, respectively. The models have been calculated with $Z=0.01$ and OPAL opacity tables. The grey area depicts the error of $\log T_\mathrm{eff}$ for HD 163899 (Saio et al. 2006).}
\label{fig8}
\end{center}
\end{figure}

\begin{figure}
\begin{center}
 \includegraphics[clip,height=88mm,angle=-90]{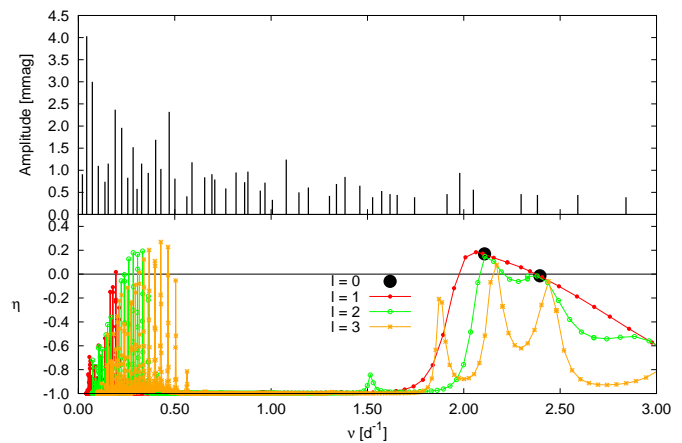}
   \caption{Comparison of frequencies detected in HD 163899 (the top panel, c.f. Table 1 in Saio et al. 2006) and theoretical frequencies from model with $M = 16 M_\odot$, $\log T_\mathrm{eff} = 4.350$ and $\log L/L_\odot = 4.701$ calculated with the OP opacities for $Z = 0.02$ (the bottom panel). Modes with the degree up to $\ell=3$ were considered.}
\label{fig9}
\end{center}
\end{figure}

The greatest uncertainty concerning SPBsg stars is their evolutionary status. Although the basic parameters ($\log T_\mathrm{eff}$, $\log g$, $Z$, $V \sin i$, etc.) of HD 163899 are very poorly determined or unknown, our calculations still can give some insights about the star. In Fig.\,\ref{fig8}, we present a comparison of evolution times in the pulsational unstable region of HR diagram during shell hydrogen burning phase and on the blue loop, for models with masses $16.0 M_\odot$ and $14.6M_\odot$, respectively, and $Z=0.01$. The star spends only a very short time (about 8 000 years) in the instability strip during the shell-hydrogen burning phase compared to the time spent on the unstable region of the blue loop (about 800 000 years). There should be much more stars in the region of HR diagram occupied by blue supergiants that undergo core helium burning than stars that moves towards red giant branch (RGB). On the one hand, this is the main and only argument why SPBsg stars should be on the blue loop. On the other hand, HD 163899 is the only know star of this type and we might be very lucky to catch it during the very short phase of pulsational instability before core helium ignition. Other arguments also prefer this evolutionary stage. The instability area during a crossing towards RGB is wider in $T_\mathrm{eff}$ than on the blue loop. This is especially important when we take into account the effect of mass loss which is ubiquitous effect for massive stars. In such a case the instability area on the blue loops become very small. The most important argument is a number of excited modes; there is much more unstable modes in models during the shell-hydrogen burning phase than in models on the blue loop. The last argument against the evolution on the blue loop is a complete lack of unstable high-frequency modes in this phase. Such modes are present in the spectrum of HD 163899 and they are present in our models before core helium ignition.

In Fig.\,\ref{fig9}, we compare the oscillation spectrum of HD 163899 (top panel) and the theoretical one computed for a model in the shell-hydrogen burning phase ($\eta$ vs. $\nu$, bottom panel). Pulsational modes with the degree up to $\ell=3$ were considered.
This is one of models in which the larger number of pulsation modes are excited. The observed spectrum consists of 48 frequencies between $0.02$ to $2.85$ c/d (c.f. Table 1 in Saio et al. 2006). This range covers both g-modes (high-order) and p-modes in post-MS models. There are also frequencies between those two groups which do not occur in pulsation calculations. We do not know $V \sin i$ for the star and hence we do not include rotation in our calculations. This is left for the future work. The estimated critical velocity for Model\,1, which has the stellar radius $R=15.5~R_\odot$, is $V_\mathrm{crit}\simeq 440$ km/s which corresponds to the critical rotation frequency of about $\nu_\mathrm{crit}\simeq 0.56$ c/d. It is very unlikely that a supergiant would rotate at such a high rate, but even with lower rotation rate the very low dominant frequency, $\nu_1 \approx 0.04$ c/d, could be explained by retrograde modes. The rotational splitting could also solve the biggest problem of our pulsation models which is too low number of unstable g-modes. As for the observed range of frequencies which occur between the theoretical g- and p-mode maxima,
it could be filled in with the high $\ell$ modes of various azimuthal orders, $m$.

\section{Conclusions}

Our goal was to study pulsational properties of B-type supergiants with masses $M<20M_\odot$, especially to check the possibility whether SPBsg stars can undergo core helium burning. In addition to previously known SPBsg instability strip (Saio et al. 2006, D2013) which covers the stars that move towards RGB after departure from MS, we found the instability strip for the blue-loop models for the temperatures of B spectral types. Its width depends on the extension and shape of the blue loops and hence it varies with many parameters, especially with metallicity, overshooting and mass loss.

One of the most important results of our paper is the fact that an inward overshooting from the outer convective zone on the RGB phase
is indispensable for the emergence of blue loops at all.
The lower values of metallicity in the main sequence and shell hydrogen burning models acts towards decreasing pulsational instability whereas
in the case of the blue loops models the most important factor is the effective temperature.
For lower values of $Z$ the blue loops have larger extension and the driving layer is located at the proper place for
pulsation to be excited. We found unstable modes in the blue loop models with metallicity  as low as $Z\approx0.004$.

There is a difference in behaviour of instability parameter, $\eta$, between models before and after core helium ignition. As we previously shown (D2013), in models during shell-hydrogen burning phase there are two groups of unstable mods; one related to the very high-order g-modes and the second related to the p-mode behaviour. On the contrary, in models on the blue loops only the very high-order g-modes are unstable and the number of unstable modes is much lower than in models that departed from TAMS and move towards RGB. That might be explained by the difference in the kinetic energy density distribution between those two evolutionary phases. In models that undergo shell hydrogen burning, only a partial reflection of pulsation mode at the ICZ is sufficient to efficiently drive the mode, whereas on the blue loop the mode has to be almost entirely reflected in order to be unstable.

Comparing the properties of models before and after core helium ignition and frequencies detected in the light curve of HD 163899, we drew a conclusion that SPBsg stars should rather undergo shell hydrogen burning and not core helium burning on the blue loop. However, to fully support this statement a better determination of basic parameters of HD 163899 is needed and, in the next step, seismic model of the stars should be constructed taking into account the effects of rotation.

We found also that mode identification from multicolour photometry is applicable to SPBsg stars and hence multi-colour photometry for HD 163899 would be desirable. The analysis of such data could finally lead to constraining the evolutionary status of the star and help to understand pulsations of SPBsg stars.


\section*{Acknowledgments}

This work was financially supported by the Polish NCN grants 2013/09/N/ST9/00611, 2011/01/M/ST9/05914 and 2011/01/B/ST9/05448.

\end{document}